\documentclass[3p, twocolumn]{elsarticle}
\biboptions{sort&compress}
\usepackage[breaklinks=true]{hyperref}
\usepackage[anythingbreaks]{breakurl}
\usepackage{amsthm}
\usepackage{amsmath}
\usepackage{amsfonts}
\usepackage{amssymb}
\usepackage{amsfonts}
\usepackage{amsthm}
\usepackage{tikz}
\usetikzlibrary{shapes.geometric, arrows.meta}
\newtheorem{theorem}{Theorem}
\usepackage{blkarray}
\usepackage{enumitem}
\usepackage[capitalize]{cleveref}
\crefname{appendix}{}{}
\Crefname{appendix}{}{}
\usepackage[super]{nth}
\usepackage{float}
\usepackage{caption}
\usepackage{subcaption}
\usepackage{dblfloatfix}
\usepackage{xcolor}
\usepackage{graphicx, import, color}
\graphicspath{{./Images/5/}}

\usetikzlibrary{arrows.meta, 
	bending, 
	calc, chains, 
	decorations.pathmorphing,
	positioning
}

\usepackage{pgfplots}

\usepgfplotslibrary{fillbetween}
\usetikzlibrary{intersections}

\begin{document}

\begin{frontmatter}
	\title{A Quantum Linear Systems Pathway for Solving Differential Equations}
	    
    \author[a,b]{Abhishek Setty\corref{mycorrespondingauthor}}
    \cortext[mycorrespondingauthor]{Corresponding author}
    \ead{a.setty@fz-juelich.de}  
    
    \address[a]{Forschungszentrum Jülich, Institute of Quantum Control (PGI-8), D-52425 Jülich, Germany}
    \address[b]{Institute for Theoretical Physics, University of Cologne, D-50937 Cologne, Germany}
    
\begin{abstract}
We present a systematic pathway for solving differential equations within the quantum linear systems framework by combining block encoding with Quantum Singular Value Transformation (QSVT). The approach is demonstrated on a complex tridiagonal linear system and extended to problems in computational fluid dynamics: the heat equation with mixed boundary conditions and Carleman-linearized nonlinear Burgers’ equation. Our scaling analysis of the heat equation identifies regimes where classical computation remains feasible and estimates circuit depths required to achieve potential quantum advantage. We further evaluate post-selection success probabilities for the presented examples and provide hardware resource estimates for block encoding and QSVT circuits in terms of two-qubit gate depth, evaluated on IBM superconducting processors with heavy-hex and square lattice topologies. These results highlight both the practical limitations of current hardware and key directions for depth reduction and scalable quantum linear solvers.
\end{abstract}

	\begin{keyword}
		Differential equations \sep Quantum singular value transformation \sep Block encoding \sep Quantum linear algebra 
	\end{keyword}
\end{frontmatter}
\thispagestyle{empty}
\tableofcontents

\section{Introduction}\label{sec:Intro}
The quest to harness quantum algorithms for solving differential equations has been underway for over a decade. Between linear algebra \cite{harrow2009quantum,childs2017quantum} and variational \cite{setty2025self,lubasch2020variational} approaches, the demand for provably accurate and deterministic solutions has led to growing interest in linear algebra methods, which circumvent the major challenges of barren plateaus and optimization difficulties inherent in variational schemes \cite{cerezo2021cost,larocca2025barren}.

A pioneering linear-algebra-based approach is the Harrow-Hassidim-Lloyd (HHL) algorithm \cite{harrow2009quantum}, which provides a quantum procedure for solving linear systems of equations. Several works have demonstrated that linear differential equations can be reduced to linear systems via spatial or temporal discretization schemes and subsequently solved using quantum linear system solvers \cite{clader2013preconditioned,cao2013quantum,berry2014high,berry2017quantum}. The HHL algorithm employs quantum phase estimation as a subroutine to extract eigenvalue information and apply controlled transformations that implement matrix inversion. More recently, Quantum Singular Value Transformation (QSVT) has emerged as a unifying framework encompassing matrix inversion, phase estimation, Hamiltonian simulation, and amplitude amplification \cite{gilyen2019quantum, martyn2021grand}. In this framework, solving linear systems amounts to approximating $f(x) = 1/x$ through polynomial transformations of the singular values. For a matrix $A \in \mathcal{C}^{N \times N}$ with condition number $\kappa$, QSVT achieves query complexity $\mathcal{O}[\kappa \log{(\kappa/\epsilon)}]$, improving on the HHL scaling of $\mathcal{O}[\kappa^2 \log{(N)/\epsilon}]$, where $\epsilon$ denotes the error tolerance \cite{martyn2021grand}. While this complexity is independent of $N$ at the level of oracle queries, practical circuit resources may still scale with $N$ through the construction of block-encoding oracles and data-loading registers, which we explicitly analyze in this work. Although these advances are well understood theoretically, detailed circuit-level simulation, scalability analysis, and quantitative hardware resource estimation for solving differential equations remain underexplored.

In this work, we consolidate key algorithmic and implementation developments and present a systematic pathway for applying QSVT-based linear solvers to differential equations. We first provide a brief overview of matrix inversion via QSVT with gate-level block encoding as a subroutine, followed by its application to complex linear systems. We then investigate differential equations from computational fluid dynamics (CFD), specifically the linear heat equation and the Carleman-linearized nonlinear Burgers’ equation. Our scaling analysis identifies regimes where classical computation remains feasible and estimates circuit depths required to achieve potential quantum advantage. We further evaluate post-selection success probabilities in the presented examples. Finally, we provide resource estimates for block encoding and QSVT circuits in terms of two-qubit gate depth, evaluated on IBM superconducting processors with heavy-hex and square lattice topologies.

\begin{figure*}[t]
	\centering
	\begin{tikzpicture}[node distance=1.5cm]
		\tikzstyle{rounded-block} = [rectangle, rounded corners, minimum width=3.5cm, 
		minimum height=1cm,text centered, draw=black]
		
		\tikzstyle{sharp-block} = [rectangle, minimum width=3.5cm, minimum height=1cm, 
		text centered, draw=black]
		
		\tikzstyle{decision} = [diamond, aspect=2, text centered, draw=black]
		
		\tikzstyle{arrow} = [very thick,->,>=stealth]
		\node (p1) [rounded-block] {Differential equations};
		\node (p2) [sharp-block, below of=p1] {Discretization (FDM, FEM, FVM) \cite{leveque1992numerical,leveque2007finite,strikwerda2004finite,zienkiewicz2005finite,brenner2008mathematical,leveque2002finite,morton2005numerical,patankar2018numerical}};
		\node (p3) [decision, below of=p2] {Linear?};
		\node (p4) [sharp-block, right of=p3, xshift=4cm] {Carleman linearization \cite{carleman1932application,kowalski1991nonlinear,liu2021efficient,bakker2023quantum,akiba2023carleman,lewis2024limitations,wu2025quantum}};
		\node (p5) [sharp-block, below of=p4] {Determine truncation order \cite{forets2017explicit,amini2022carleman,gonzalez2025quantum}};
		\node (p6) [sharp-block, below of=p3] {Formulate $Ax = b$};
		\node (p7) [sharp-block, below of=p6, xshift=7cm] {Estimate $\sigma_{\text{min}}$ \cite{strikwerda2004finite,leveque2007finite,morton2005numerical,trefethen2022numerical,varga2011gervsgorin,golub2013matrix,saad2011numerical}};
		\node (p8) [sharp-block, below of=p7, xshift=0cm] {Generate $\phi$ for matrix inversion \cite{chao2020finding,dong2021efficient,martyn2021grand}};
		\node (p9) [sharp-block, below of=p6, yshift=-1cm] {Block encoding of sparse matrices \cite{setty2025block}};
		\node (p10) [sharp-block, below of=p9] {QSVT \cite{gilyen2019quantum}};
		\node (p11) [sharp-block, below of=p10, xshift=-3cm, yshift=-0.5cm] {State tomography \cite{cramer2010efficient}};
		\node (p12) [sharp-block, below of=p10, xshift=1.2cm, yshift=-0.5cm] {Sampling $|x\rangle$ \cite{gilyen2019quantum,aaronson2015read}};
		\node (p13) [sharp-block, below of=p10, xshift=6.5cm, yshift=-0.5cm] {Observable estimation $\langle x| M | x \rangle$ \cite{harrow2009quantum,childs2017quantum}};
		
		\draw [arrow] (p1) -- (p2);
		\draw [arrow] (p2) -- (p3);
		\draw [arrow] (p3) -- (p4) node[midway, above]{No};
		\draw [arrow] (p3) -- (p6) node[midway, left]{Yes};
		\draw [arrow] (p4) -- (p5);
		\draw [arrow] (p5) -- (p6);
		\draw [arrow] (p6.south) -- ++ (0, -1) -- (p7);
		\draw [arrow] (p6) -- (p9);
		\draw [arrow] (p7) -- (p8);
		\draw [arrow] (p8.south) -- ++ (0, -0.5) -- (p10);
		\draw [arrow] (p9) -- (p10);
		\draw [arrow] (p10.south) -- ++ (0, -0.5) -- ++ (-3, 0) -- (p11.north);
		\draw [arrow] (p10.south) -- ++ (0, -0.5) -- ++ (1.2, 0) -- (p12.north);
		\draw [arrow] (p10.south) -- ++ (0, -0.5) -- ++ (6.5, 0) -- (p13.north);
		
	\end{tikzpicture}
	\caption{Quantum linear systems pathway for solving differential equations using QSVT. Here $\sigma_{\text{min}}$ denotes the smallest singular value of the matrix, and $\phi$ denotes the phases used for approximating the inverse function (see \cref{sec:qsp,sec:inverse_function}).}
	\label{fig:1_Pathway}
\end{figure*}

\begin{figure*}[t]
	\centering
	\includegraphics[width=\textwidth]{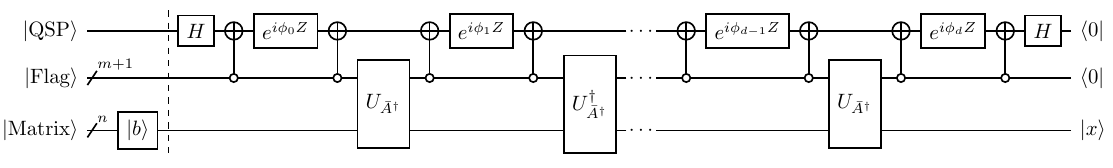}
	\caption{Quantum circuit for solving the linear system $|x\rangle = \text{QSVT}(\bar{A}^{\dag})|b\rangle$ using QSVT, with the rescaled solution given by \cref{eq:LSE}. For demonstration, a square matrix $\bar{A}$ is used, with phases $\{\phi_i\}_{i=0}^d$ in the reflection convention \cref{sec:reflection_convention} for an odd-degree $d$ inverse polynomial (see \cref{sec:inverse_function,eq:U_phi_QSVT}). The $|\text{QSP}\rangle$ qubit applies the phase sequence $\vec{\phi}$. The $|\text{Flag}\rangle$ register contains $m$ data qubits plus one delete qubit as in \cref{fig:A3_Block_Encode}. The $|\text{Matrix}\rangle$ register consists of $n$ qubits for preparing $|b\rangle$ and encoding $\bar{A}^{\dag}$. All qubits start in $|0\rangle$ state. Projector-controlled phase shift gates $\raisebox{0.3ex}{\scalebox{0.7}{$\prod$}}_{\phi_i}$ and $\raisebox{0.3ex}{\scalebox{0.7}{$\widetilde{\prod}$}}_{\phi_i}$ (identical for a square matrix) are implemented using MCX gates with controls on all $|\text{Flag}\rangle = 0$. The vertical dotted line indicates a barrier separating $|b\rangle$ from the rest of the circuit; the dots indicate a repeated pattern. Post-selection on the $|\text{QSP}\rangle$ and $|\text{Flag}\rangle$ is denoted by $\langle 0|$, and the final solution state is $|x\rangle$.}
	\label{fig:2_QSVT}
\end{figure*}

\section{A Linear Systems Pathway}\label{sec:pathway}
In this section, we outline a pathway for solving differential equations through their reduction to linear systems of equations \cref{fig:1_Pathway}. Differential equations that admit discretization via finite difference (FDM), finite element (FEM), or finite volume (FVM) methods yield systems of the form $Ax = b$ when linear. For nonlinear cases, Carleman linearization is applied, with the truncation order chosen based on the degree of nonlinearity and desired error tolerance, again resulting in a system $Ax=b$.

The smallest singular value $\sigma_{\min}$ of $A$ is estimated using spectral properties of the discretization matrix. For near-Toeplitz matrices arising from finite difference operators, analytical eigenvalue expressions provide accurate estimates of extremal singular values \cite{strikwerda2004finite,leveque2007finite,morton2005numerical}. For structured but non-Toeplitz matrices, such as those obtained from Carleman linearization, spectral bounds are obtained using perturbation arguments relative to the underlying linear operator \cite{trefethen2022numerical}, Gershgorin-type eigenvalue bounds \cite{varga2011gervsgorin}, or classical iterative methods such as power iteration and Lanczos algorithms \cite{golub2013matrix,saad2011numerical}. These approaches provide reliable estimates from which the factor $\kappa$ is determined. This value informs the construction of phase factors $\phi$ that approximate the inverse function polynomial using quantum signal processing (refer \cref{sec:qsp,sec:inverse_function}). The matrix is subsequently block-encoded, and the inversion phases $\phi$ are applied within the QSVT framework to solve the linear system.

The procedure outputs a quantum state $|x\rangle$, from which information can be extracted in several ways: (i) full tomography, which sacrifices exponential speedup, (ii) sampling to probe the probability distribution and identify dominant components, or (iii) computing observable expectation values $\langle x|M|x \rangle$ for overlaps or other quantities of interest.

\section{Matrix Inversion by QSVT}\label{sec:matrix_inversion_by_qsvt}
In this section, we elaborate on the procedure for matrix inversion using QSVT, building on the concepts described in \cref{sec:qsp,sec:inverse_function,sec:block_encoding,sec:qsvt}. Starting from a linear system $A\vec{x} = \vec{b}$, the goal is to solve for $\vec{x}$ as $\vec{x} = A^{+}\vec{b}$, where $A^{+}$ denotes the pseudoinverse of $A$. 

QSVT relies on block encoding, which embeds a (possibly non-unitary) matrix $A$ into a unitary $U_A$ as,
\begin{equation}\label{eq:U_A_Block_Encode}
	U_A = 
	\begin{pmatrix}
		A/\alpha & *\\
		* & *
	\end{pmatrix},
\end{equation}
where $||\cdot||_2$ is the spectral norm and $\alpha \geq ||A||_2$ is a subnormalization factor which guarantees that the singular values of $A/\alpha$ lie in $[0,1]$ for the QSVT polynomial (see \cref{theorem_qsp}).

Let the singular value decomposition (SVD) of $A$ \cref{eq:A_svd} be
\begin{equation}
 	\bar{A} = A/\alpha = \sum_k \sigma_k |w_k\rangle \langle v_k |.
 \end{equation}
 The pseudoinverse is then given by
 \begin{equation}
 	\bar{A}^{+} = \alpha A^{+} = \sum_k \frac{1}{\sigma_k} |v_k\rangle \langle w_k |.
 \end{equation}
By block encoding the conjugate transpose $\bar{A}$ as $\bar{A}^{\dag} = A^{\dag}/\alpha = \sum_k \sigma_k |v_k\rangle \langle w_k |$, the pseudoinverse can be approximated using QSVT:
 \begin{equation}
 	\bar{A}^{+} \approx \text{QSVT}(\bar{A}^{\dag}) = \sum_k \text{Poly}(\sigma_k) |v_k\rangle \langle w_k |,
 \end{equation}
 where the QSVT construction is described in \cref{sec:qsvt}. 
 
Representing $\vec{b} \in \mathbb{C}^N$ as a quantum state $|b\rangle = \vec{b}/||\vec{b}||$, the solution to the linear system is
 \begin{equation} \label{eq:LSE}
 	\vec{x} = A^{+}\vec{b} \approx \text{QSVT}(\bar{A}^{\dag}) |b\rangle \left( \frac{||\vec{b}||}{\alpha \beta} \right),
 \end{equation}
 where $\beta$ is a scaling factor (see \cref{eq:beta}).
 
The quantum circuit implementing QSVT is shown in \cref{fig:2_QSVT}, where the desired state $|x\rangle = \text{QSVT}(\bar{A}^{\dagger})|b\rangle$ is obtained upon successful post-selection with probability $p_s \in [0, 1]$. The full output state of the circuit can be written as
\begin{equation}
	\resizebox{\columnwidth}{!}{$
	|\psi_{\text{QSVT}}\rangle = |0\rangle_{\text{QSP}} \otimes |0^{\otimes m+1}\rangle_{\text{Flag}} \otimes |x\rangle + |*\rangle,$}
\end{equation}
 where $|\psi_{\text{QSVT}}\rangle$ is the output state of the circuit, $|*\rangle$ denotes an orthogonal junk component corresponding to all other measurement outcomes. The probability of successfully measuring the $|\text{QSP}\rangle$ and $|\text{Flag}\rangle$ qubits in the $|0\rangle$ state (post-selection) is given by the squared norm of the desired component,
 \begin{equation}\label{eq:p_success_analytical}
 	\begin{split}
 		 	p_s &= || \; |x\rangle \; ||^2 = || \text{QSVT}(\bar{A}^{\dagger}) | b\rangle || ^2, \\
 		 	&= || \sum_k \text{Poly}(\sigma_k)|v_k\rangle\langle w_k| \sum_i \langle w_i |b\rangle |w_i\rangle ||^2,\\
 		 	& \approx || \sum_k \beta \frac{1}{\sigma_k} \langle w_k|b\rangle |v_k\rangle ||^2,\\
 		 	& = \sum_k \left( \frac{\beta}{\sigma_k} \right)^2 |\langle w_k | b\rangle|^2.
 	\end{split}
 \end{equation}
 
Alternatively, the success probability can be estimated directly from measurement outcomes of $|\text{QSP}\rangle$ and $|\text{Flag}\rangle$ qubits,
\begin{equation}\label{eq:p_success}
	p_s = \frac{N_{\text{success}}}{N_{\text{total}}},
\end{equation}
where $N_{\text{total}}$ denotes the total number of measurement shots and $N_{\text{success}}$ counts the shots where the $|\text{QSP}\rangle$ and $|\text{Flag}\rangle$ qubits are measured in the $|0\rangle$ state.
 
Note that the circuit shown in \cref{fig:2_QSVT} is for square matrices. For non-square matrices, the projectors $\raisebox{0.3ex}{\scalebox{0.7}{$\prod$}}$ and $\raisebox{0.3ex}{\scalebox{0.7}{$\widetilde{\prod}$}}$ can be adjusted to select the appropriate left and right singular vectors. Block-encoded matrices $U_A$ are of size $2^n \times 2^n$, where $n$ is the number of qubits. If the matrix to be inverted does not occupy the full space, the desired submatrix $B_1$ can be isolated from $\begin{bmatrix}
	B_1 & B_2 \\
	B_3 & B_4
\end{bmatrix}$ either by choosing suitable projectors within QSVT or by setting $B_2=0$ and $B_3=0$.

 \section{Circuit Level Simulations}\label{sec:Circuit_level_simulations}
We perform circuit-level simulations to evaluate QSVT-based matrix inversion for linear systems, starting with a complex tridiagonal system. Applications include differential equations from CFD, such as the heat equation with Dirichlet and Neumann boundary conditions and the nonlinear Burgers’ equation via Carleman linearization. Simulations are carried out using PennyLane \cite{bergholm2018pennylane}, while hardware resource estimates are evaluated for current IBM superconducting quantum processors.

\begin{figure*}[t]
	\centering
	\begin{subfigure}[t]{0.4\textwidth}
		\includegraphics[width=\textwidth]{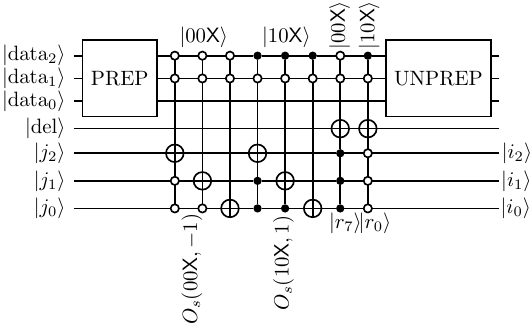}
		\caption{} 
		\label{fig:3a_Complex}
	\end{subfigure}
	\hfill
	\begin{subfigure}[t]{0.29\textwidth}
		\includegraphics[width=\textwidth]{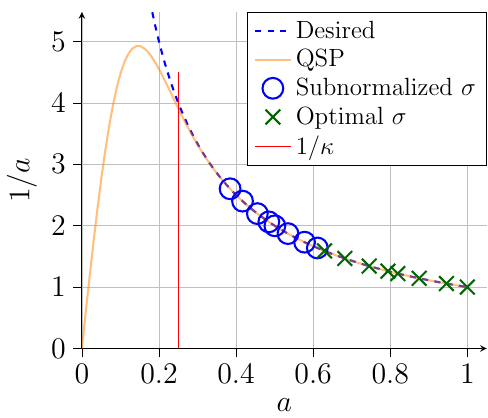}
		\caption{}
		\label{fig:3b_Complex}
	\end{subfigure}
	\hfill
	\begin{subfigure}[t]{0.29\textwidth}
		\includegraphics[width=\textwidth]{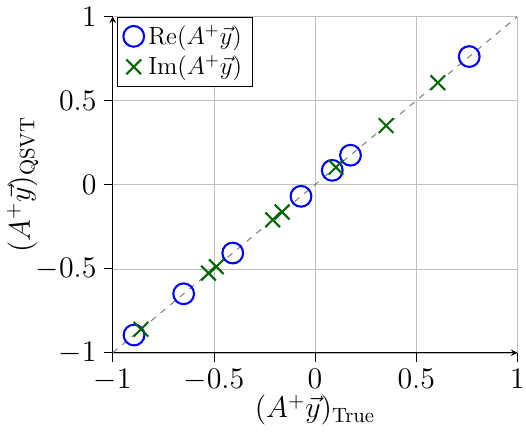}
		\caption{}
		\label{fig:3c_Complex}
	\end{subfigure}	
	\caption{Solution of the complex tridiagonal linear system \cref{sec:tridiagonal_complex}. 
		\subref{fig:3a_Complex} Circuit for block encoding $A^{\dagger}$ of \cref{eq:complex_matrix}. \subref{fig:3b_Complex} Inverse function approximation using Quantum Signal Processing (QSP) \cref{sec:qsp,sec:inverse_function}. Subnormalized $\sigma$ denote singular values of $A^{\dagger}/\alpha$, while optimal $\sigma$ denote singular values of $A^{\dagger}/||A^{\dagger}||_2$, with $||\cdot||_2$ the spectral norm. 
		\subref{fig:3c_Complex} Parity plot comparing the true $(A^{+}\vec{y})_{\text{True}}$ with the QSVT solution $(A^{+}\vec{y})_{\text{QSVT}}$, showing real and imaginary parts separately.}
	\label{fig:3_Tridiagonal_complex}
\end{figure*}
 
\subsection{Complex Tridiagonal Linear System}\label{sec:tridiagonal_complex}
Consider the linear system $A\vec{x} =\vec{y}$, where $A \in \mathbb{C}^{8 \times 8}$ is a complex tridiagonal coefficient matrix and $\vec{y} \in \mathbb{C}^8$ is a complex vector:
\begin{equation}\label{eq:complex_matrix}
	A = 
\begin{bmatrix}
	z_2 & z_3\\
	z_1 & z_2 & z_3\\
	& \ddots & \ddots & \ddots\\
	&  & z_1 &z_2 & z_3\\
	& & &z_1 & z_2
\end{bmatrix},
\end{equation}
where $z_1 = \psi_0 + \psi_1\textrm{i}, z_2 = \psi_2 + \psi_3\textrm{i}, z_3 = \psi_4+\psi_5\textrm{i}$ and $\{\psi_i\}_{i=0}^{5} \in (-1, 1)$ are sampled independently from a uniform distribution. Similarly, each component of $\vec{y}$ is generated as $y_k = \gamma_k + \delta_k\textrm{i}$ with $\gamma_k, \delta_k, \sim \mathcal{U}(-1, 1)$. 

\begin{figure*}[t]
	\centering
	\begin{subfigure}[t]{0.4\textwidth}
		\centering
		\includegraphics[width=\textwidth]{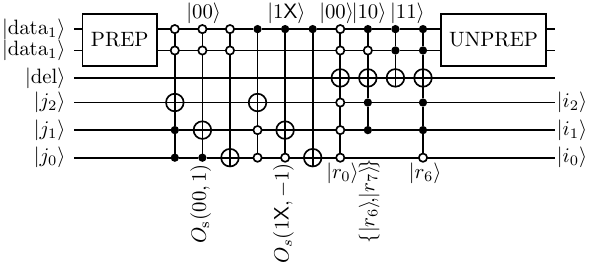}
		\caption{} 
		\label{fig:4a_Heat}
	\end{subfigure}
	\hfill
	\begin{subfigure}[t]{0.2935\textwidth}
		\centering
		\includegraphics[width=\textwidth]{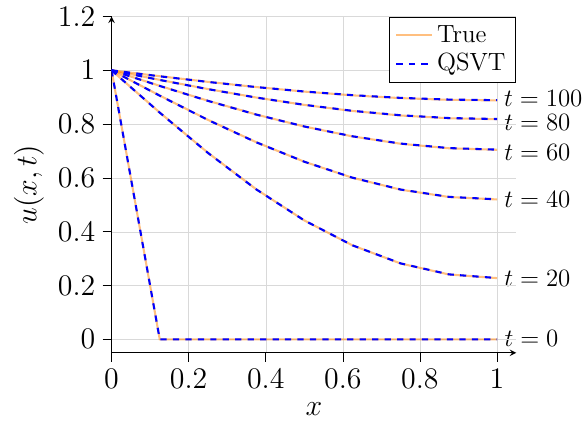}
		\caption{}
		\label{fig:4b_Heat}
	\end{subfigure}
	\hfill
	\begin{subfigure}[t]{0.2935\textwidth}
		\centering
		\includegraphics[width=\textwidth]{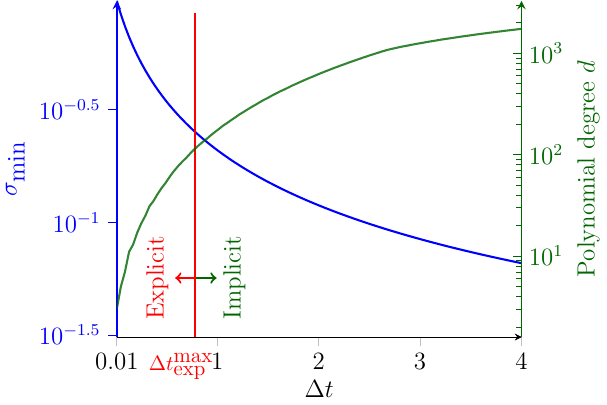}
		\caption{}
		\label{fig:4c_Heat}
	\end{subfigure}	
	\\
	\begin{subfigure}[t]{0.5\textwidth}
		\centering
		\includegraphics[width=\textwidth]{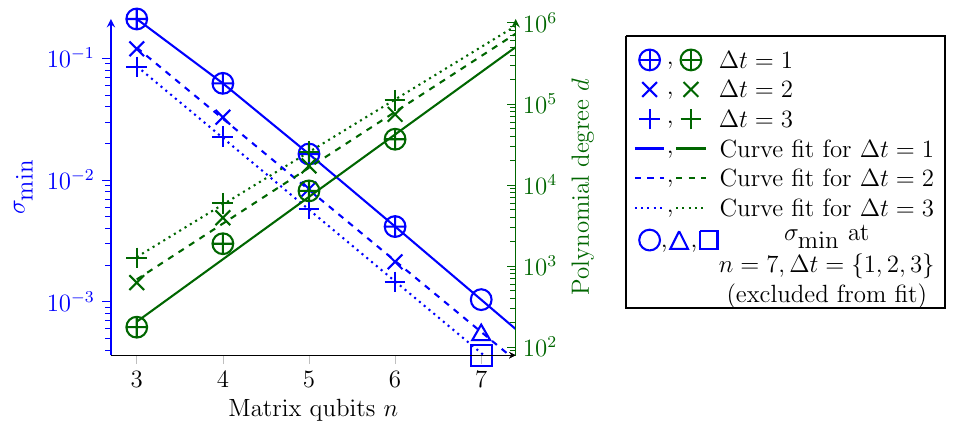}
		\caption{}
		\label{fig:4d_Heat}
	\end{subfigure}	
	\hspace{0.5cm}
	\begin{subfigure}[t]{0.33\textwidth}
		\centering
		\includegraphics[width=\textwidth]{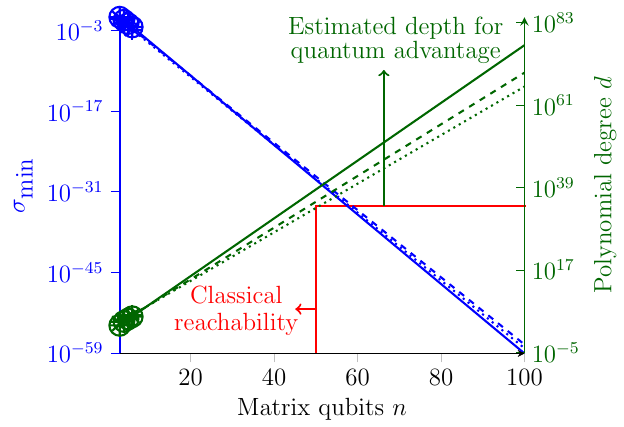}
		\caption{}
		\label{fig:4e_Heat}
	\end{subfigure}	
	\caption{Solution of the heat equation \cref{sec:heat_equation}. \subref{fig:4a_Heat} Circuit for block encoding $A^{\dag}$ of matrix \cref{eq:heat_A_numerical}. \subref{fig:4b_Heat} Implicit solution with left Dirichlet and right Neumann boundary conditions via QSVT, compared with the true solution, for $t=[0, 100]$. \subref{fig:4c_Heat} Scaling of the minimum singular value $(\sigma_{\text{min}})$ and inverse polynomial degree $d$ versus time step $\Delta t (\Delta x = 0.125)$; $\Delta t_{\text{exp}}^{\text{max}}$ indicates the threshold separating explicit and implicit schemes. \subref{fig:4d_Heat} Scaling with the number of matrix qubits $n=\{3, 4, 5, 6\}$ for $\Delta t = \{1, 2, 3\}$. A double exponential fit is shown and tested for $n = 7$. \subref{fig:4e_Heat} Estimated $\sigma_{\text{min}}$ and polynomial degree as $n$ scales, with the same legend as \subref{fig:4d_Heat}. Extrapolation can estimate circuit depth required for quantum advantage over a 50-qubit classical simulator.}
	\label{fig:4_Heat}
\end{figure*}

The solution procedure via QSVT follows \cref{sec:matrix_inversion_by_qsvt}, with the block encoding applied to $A^{\dag}$. Using the block encoding protocol \cite{setty2025block}, we create the data vector $v_{\text{data}} = [|\psi_0|, |\psi_1|, |\psi_2|, |\psi_3|, |\psi_4|, |\psi_5|]^T$ and sign vector $v_{\text{sign}} = [\text{sgn}(\psi_0), -\text{sgn}(\psi_1)\textrm{i}, \text{sgn}(\psi_2), -\text{sgn}(\psi_3)\textrm{i},\\ \text{sgn}(\psi_4), -\text{sgn}(\psi_5)\textrm{i}]^T$, where $\text{sgn}(a)=\frac{a}{|a|}$ is a sign function. We embed the data (refer \cite{mottonen2004transformation,mottonen2004quantum}) into three data qubits $(|\text{data}\rangle)$. Data elements ${\{\psi_0, -\psi_1\textrm{i}\}}$ in basis states $\{|000\rangle, |001\rangle\}$ (compressed to $|00\mathsf{X}\rangle$) should be shifted right by one column $O_s(k=00\mathsf{X}, diag=-1)$ (see \cref{eq:O_shift}) and deleted in row $|r_7\rangle$. Here $\mathsf{X}$ denote no control on $|\text{data}_0\rangle$ qubit. Similarly, data elements ${\{\psi_4, -\psi_5\textrm{i}\}}$ in basis states $\{|100\rangle, |101\rangle\}$ (compressed to $|10\mathsf{X}\rangle$) should be shifted left by one column $O_s(k=10\mathsf{X}, diag=1)$ (see \cref{eq:O_shift}) and deleted in row $|r_0\rangle$. The circuit for this matrix is shown in \cref{fig:3a_Complex}.

QSVT applies a polynomial function to the singular values of the block encoded matrix $A^{\dag}/\alpha$ without explicitly extracting them. To demonstrate inversion, we implemented quantum signal processing for the inverse function \cref{fig:3b_Complex}, following the pathway in \cref{sec:pathway}. The minimum singular value $\sigma_{\text{min}}$ is estimated, determining the factor $\kappa = 4$ resulting in polynomial degree $d=117$ for the inverse approximation. Subnormalization of $A^{\dag}$ reduces singular values, which increases the required polynomial degree; note that $\sigma_{\text{min}} > 1/\kappa$ is required for accurate inversion.

State-vector simulations were performed to replicate the effect of full state tomography. The solution $\vec{x} =A^{+} \vec{y}$ is shown in \cref{fig:3c_Complex}, with real and imaginary parts plotted separately in a parity plot. These results demonstrate that sparse complex linear systems can be solved effectively using block encoding \cite{setty2025block} and QSVT \cite{gilyen2019quantum}, as proposed in this work.

The hardware resources are estimated for current IBM superconducting quantum processors with heavy-hex and square lattice topologies and are presented in \cref{table:Resource_1}. The resources are quantified in terms of two-qubit depth, defined as the transpiled circuit depth restricted to native two-qubit operations after mapping to device connectivity constraints. This corresponds to the critical path of sequential two-qubit gates that determines the dominant contribution to circuit runtime on superconducting hardware. We also estimate the post-selection success probability both analytically (see \cref{eq:p_success_analytical}) and numerically (see \cref{eq:p_success}), and report the results in \cref{table:Resource_1}.

\begin{table}[ht]
	\centering
	\begin{tabular}{|c|c|}
		\hline
		\multicolumn{2}{|c|}{Complex tridiagonal linear system}\\
		\hline
		$\text{matrix qubits} (n) $ & $3$ \\
		$\text{total qubits}$ & $8$\\
		$\kappa$ & $4$\\
		$d$ & $117$\\
		$p_s$ (analytical) & $0.16706$\\
		$N_{\text{total}}$ & $10^9$\\
		$p_s$ (numerical) & $0.16702$\\
		$\text{BE}_{\text{Hex}}$ & $283$\\
		$\text{BE}_{\text{Square}}$ & $240$\\
		$\text{QSVT}_{\text{Hex}}$ & $57$K\\
		$\text{QSVT}_{\text{Square}}$ & $44$K\\
		\hline
	\end{tabular}
	\caption{Estimation of success probability and hardware resources, measured in terms of two-qubit depth, for the complex tridiagonal linear system described in \cref{sec:tridiagonal_complex}. Here $p_s$ (analytical) refers to \cref{eq:p_success_analytical} and $p_s$ (numerical) refers to \cref{eq:p_success}. `K' denotes thousands (e.g., 57K denotes $57 \times 10^3$ two-qubit gate depth).`BE' denotes the block encoding circuit as in \cref{fig:3a_Complex} and `QSVT' denotes the complete QSVT circuit as in \cref{fig:2_QSVT}. `Hex' denotes the heavy-hex lattice topology of the Heron r3 processor (IBM boston), and `Square' denotes the square lattice topology of the Nighthawk r1 processor (IBM miami).}
	\label{table:Resource_1}
\end{table}

\subsection{Heat Equation With Dirichlet and Neumann Boundary Conditions}\label{sec:heat_equation}
We consider the 1D heat equation $\frac{\partial u}{\partial t} = \nu \frac{\partial^2 u}{\partial x^2}, x \in [0, 1], t > 0$, with initial condition $u(x, 0)=0, x \in (0, 1]$, left Dirichlet boundary $u(0, t) = 1$, and right Neumann boundary $\frac{\partial u}{\partial x}|_{x=1} = 0$. Discretization via finite differences allows explicit (\cref{eq:heat_explicit}) and implicit (\cref{eq:heat_implicit}) methods. While the explicit scheme is conditionally stable with maximum allowable time step $\Delta t_{\text{exp}}^{\text{max}}$ given by $\Delta t_{\text{exp}}^{\text{max}} \leq \Delta x^2/2\nu$ (refer \cref{eq:heat_von_Neumann}), the implicit scheme is unconditionally stable and requires matrix inversion (\cref{eq:heat_LSE}).

For three matrix qubits $(n=3)$ and $N=9$ grid points, with $\nu=0.01, \Delta x = 0.125,$ and $\Delta t = 1$, the explicit method fails $(\Delta t > \Delta t_{\text{exp}}^{\text{max}})$. The implicit method is formulated as a linear system $A \vec{u} = \vec{b}$ with
\begin{equation}\label{eq:heat_A_numerical}
	A = \begingroup
	\setlength{\arraycolsep}{2.3pt}
	\renewcommand{\arraystretch}{1.2}
	\begin{bmatrix}
		2.28 &-0.64\\
		-0.64&2.28&-0.64\\
		&\ddots&\ddots&\ddots \\
		&&-0.64&2.28&-0.64\\
		&&&-1.28&2.28
	\end{bmatrix},
	\endgroup
\end{equation}
here the factor $\lambda = \frac{\nu \Delta t}{\Delta x^2} = 0.64$. 

\begin{table}[ht]
	\centering
	\begin{tabular}{|c|c|c|}
		\hline
		\multicolumn{3}{|c|}{Heat equation}\\
		\hline
		$\text{matrix qubits} (n)$ & $3$ & $4$\\
		$\text{total qubits}$ & $7$ & $8$\\
		$\kappa$ & $8$ & $12$\\
		$d$ & $559$ & $1275$\\
		$p_s$ (analytical) & $0.04223$ & $0.0486$\\
		$N_{\text{total}}$ & $10^9$ & $10^9$\\
		$p_s$ (numerical) & $0.04222$ & $0.0482$\\
		$\text{BE}_{\text{Hex}}$ & $311$ & $550$\\
		$\text{BE}_{\text{Square}}$ & $255$ & $396$\\
		$\text{QSVT}_{\text{Hex}}$ & $258$K & $903$K\\
		$\text{QSVT}_{\text{Square}}$ & $202$K & $696$K\\
		\hline
	\end{tabular}
	\caption{Estimation of success probability and hardware resources, measured in terms of two-qubit depth, for the heat equation described in \cref{sec:heat_equation}. For this analysis, we fixed $\nu = 0.01, \Delta t = 1$. Here $p_s$ (analytical) refers to \cref{eq:p_success_analytical} and $p_s$ (numerical) refers to \cref{eq:p_success}. `K' denotes thousands (e.g., 258K denotes $258 \times 10^3$ two-qubit gate depth). `BE' denotes the block encoding circuit as in \cref{fig:4a_Heat} and `QSVT' denotes the complete QSVT circuit as in \cref{fig:2_QSVT}. `Hex' denotes the heavy-hex lattice topology of the Heron r3 processor (IBM boston), and `Square' denotes the square lattice topology of the Nighthawk r1 processor (IBM miami).}
	\label{table:Resource_2}
\end{table}

\begin{figure*}[t]
	\centering
	\begin{subfigure}[t]{0.33\textwidth}
		\centering
		\scalebox{0.7}{
			\def\svgwidth{\textwidth}
			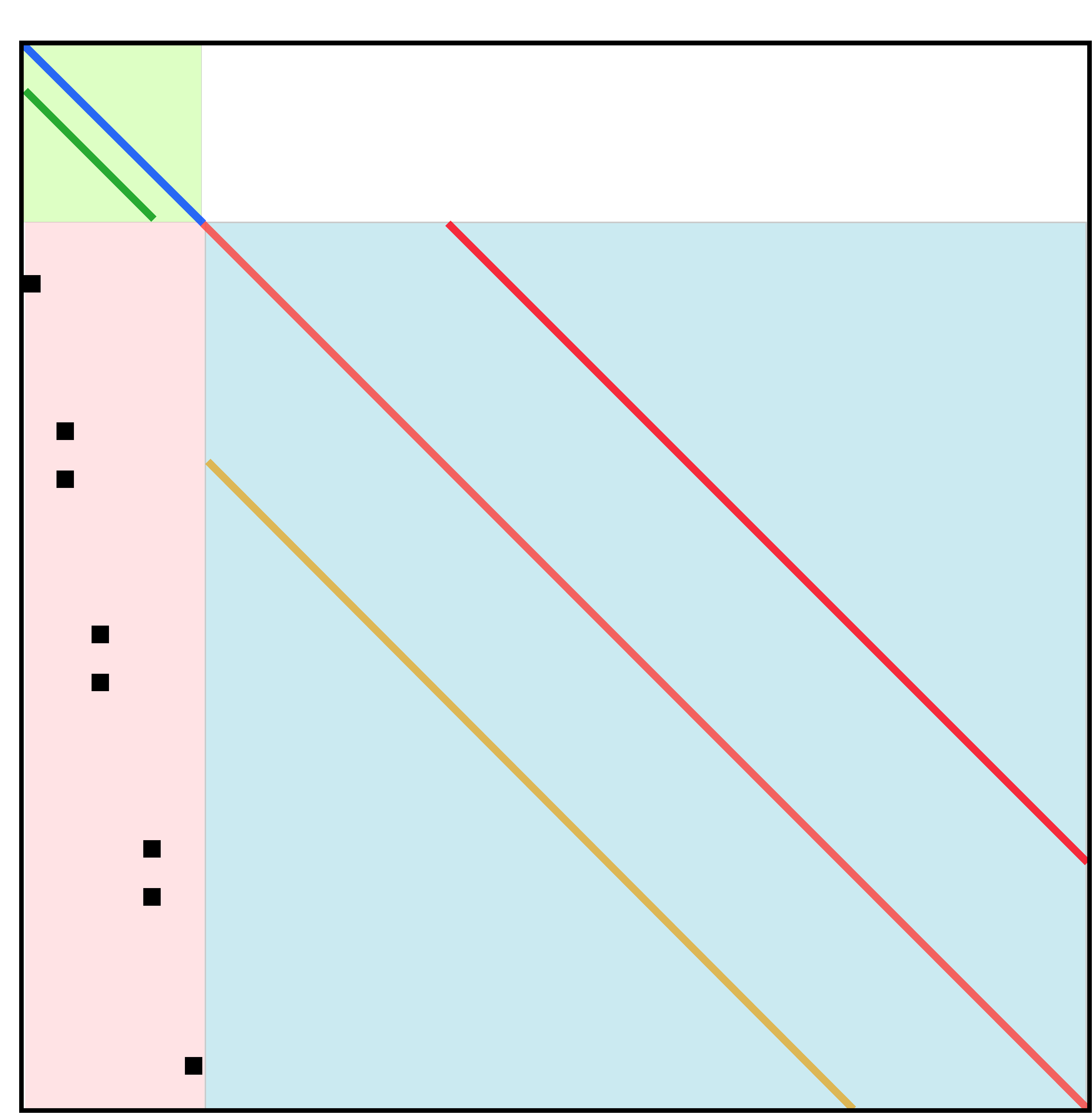
		}
		\caption{}
		\label{fig:5a_Burgers}
	\end{subfigure}
	\begin{subfigure}[t]{0.5\textwidth}
		\centering
		\includegraphics[width=\textwidth]{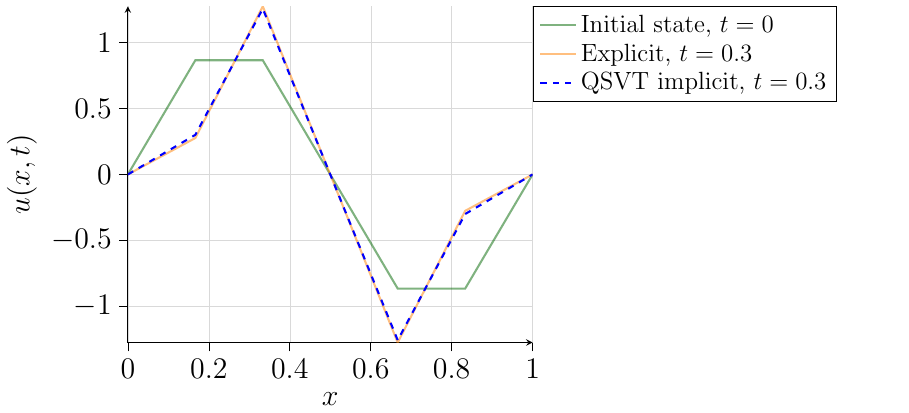}
		\caption{}
		\label{fig:5b_Burgers}
	\end{subfigure}	
	\label{fig:5_Burgers}
	\caption{\subref{fig:5a_Burgers} Matrix representation of $L^{\dagger}$ for solving Burgers’ equation. Colored sub-blocks highlight distinct matrix components, while the uncolored block corresponds to $A_1^2 = 0$ in \cref{eq:A_truncation}. Data elements are labeled $[\psi_i]_{i=0}^{13}$. Continuous colored lines indicate repeating $\psi_i$ values along diagonal $diag$ (row minus column), and black square dots mark single, non-repeating entries. \subref{fig:5b_Burgers} Initial condition at $t=0$, and comparison of the QSVT implicit solution with the classical explicit solution at $t=0.3$.}
\end{figure*}

For block encoding of matrix $A^{\dag}$ \cite{setty2025block}, the data values are $\psi_0=-0.64, \psi_1=2.28, \psi_2=-0.64, \psi_3=-1.28$, the data vector $v_{\text{data}} = [|\psi_0|, |\psi_1|, |\psi_2|, |\psi_3|]^T$ and sign vector $v_{\text{sign}}=[-1, 1, -1, -1]^T$. To construct the block encoding circuit, the data element $\psi_0$ in basis state $|00\rangle$ is shifted left by one column $O_s(k=00, diag=1)$ (see \cref{eq:O_shift}) and deleted from row $|r_0\rangle$. The data elements $\{\psi_2, \psi_3\}$ in basis states $\{|10\rangle, |11\rangle\}$ (compressed to $1\mathsf{X}\rangle$) is shifted right by one column $O_s(k=1\mathsf{X}, diag=-1)$ (see \cref{eq:O_shift}), where data element $\psi_2$ is deleted from rows $\{|r_6\rangle, |r_7\rangle\}$, and data element $\psi_3$ in basis state $|11\rangle$ is inserted into row $|r_6\rangle$. The resulting circuit implementing this transformation is shown in \cref{fig:4a_Heat}. Using the pathway in \cref{sec:pathway}, we estimate $\sigma_{\text{min}}$, set $\kappa = 8$, and calculate phases $\phi$ for an inverse polynomial of degree $d=559$. Matrix inversion via QSVT (\cref{sec:matrix_inversion_by_qsvt}) yields the solution compared with the true result in \cref{fig:4b_Heat} over $t \in [0, 100]$.

Since CFD requires fine discretization in space and time, it is crucial to analyze how quantum resource demands scale as discretization grows exponentially with the number of qubits. We study scaling with $\Delta t \in [0.01, 4]$ at fixed $\Delta x = 0.125$ (\cref{fig:4c_Heat}), showing that decreasing $\sigma_{\text{min}}$ of the subnormalized matrix and increasing $d$ flatten at larger $\Delta t$. The threshold time step $\Delta t_{\text{exp}}^{\text{max}}$ illustrates the instability of explicit schemes and the need for implicit formulations. Scaling with matrix qubits $n=\{3, 4, 5, 6\}$ and $\Delta t = \{1, 2, 3\}$ is shown in \cref{fig:4d_Heat}, where a double-exponential fit $y(x) = A e^{-Bx} + C e^{-Dx}$ \cite{virtanen2020scipy} is validated for $n=7$. Extrapolation to larger $n$ predicts the QSVT circuit depth required for quantum advantage. This extrapolation is consistent with the analytical estimation of the minimum singular value (see \cref{eq:sigma_min_estimation,eq:sigma_min_approx}), derived in \cref{sec:FDE_heat}. Since excessive circuit depth can erase speedups, even if gates are operated near the quantum speed limit \cite{margolus1998maximum,levitin2009fundamental,deffner2017quantum}, resource reduction remains essential. Moreover, this extrapolation strategy provides a benchmark to compare resource scaling of various quantum algorithms against classical reachability. The analysis of hardware resources and post-selection success probability are performed for $n = \{3, 4\}$ qubits and are reported in \cref{table:Resource_2}.

\subsection{Burgers' Equation}\label{sec:Burgers_equation}
We now consider the nonlinear Burgers' equation, $\frac{\partial u}{\partial t} + u \frac{\partial u}{\partial x} = \nu \frac{\partial^2 u}{\partial x^2}$ defined on $x \in [0, 1]$, with initial condition $u(x, 0) = \sin{(2 \pi x)}$ and Dirichlet boundary conditions $u(0, t) = u(1, t) = 0$ for $t>0$. Following the pathway in \cref{sec:pathway}, the partial differential equation (PDE) is first discretized in space via finite differences, yielding a system of nonlinear ordinary differential equations (ODEs). Carleman linearization \cite{liu2021efficient} is then applied to embed the nonlinear dynamics into a high-dimensional linear system. Time discretization using the implicit scheme produces a linear system solvable via QSVT-based matrix inversion, as detailed in \cref{sec:Carleman_Linearization,sec:Linearization_Burgers,sec:matrix_inversion_by_qsvt}.

Using the notation in \cref{sec:Carleman_Linearization,sec:Linearization_Burgers}, we consider $S=7$ grid points, $\Delta x = 1/(S-1), \nu =0.01, \Delta t = 0.1,$ total time $T = 0.3$, and truncation order $N=2$. For discrete time steps $t_k = k \Delta t$, the implicit scheme results in the linear system $L \vec{y}^{k+1} = B$ (\cref{eq:Carleman_linear_system}). Block encoding the matrix $L^{\dag}$ of size $30 \times 30$ (\cref{eq:A_truncation}) requires $5$ matrix qubits ($32 \times 32$ embedding), with the target submatrix chosen as $\begin{bmatrix}
	L^{\dag} & 0\\
	0 & *
\end{bmatrix}$, where $*$ is an arbitrary matrix (see \cref{sec:matrix_inversion_by_qsvt}). The sparse structure of $L^{\dag}$ is shown in \cref{fig:5a_Burgers}. Block encoding of $L^{\dag}$ consists of $v_{\text{data}} = [|\psi_i|]^T, i \in [0, 13]$ and sign vector is $v_{\text{sign}} = [\text{sgn}(\psi_i)]^T, i \in [0, 13]$. The number of data qubits required for block encoding is $\left\lceil \text{dim}(v_{\text{data}}) \right\rceil = 4$, where the state vector is padded with zeros for state preparation. The associated block encoding operations such as shift, deletion, and insertion are listed in \cref{table1:Burgers_operations} (see \cite{setty2025block}).

\begin{table}[ht]
	\centering
	\begin{tabular}{|p{7.2cm}|}
		\hline
		\multicolumn{1}{|c|}{Block encoding operations}\\
		\hline
		$\psi_0=-0.036, O_{\text{s}}(0000, 5), D^{|0000\rangle}_{\{0-9,30,31\}}$\\[2pt]
		$\psi_1=-0.036, O_{\text{s}}(0001, 1), D^{|0001\rangle}_{\{0,5,10,15,20,25,30,31\}}$\\
		$\tilde{\psi}_2=\psi_3-\psi_2=0.072, D^{|0010\rangle}_{\{0-4\}}$\\
		$\tilde{\psi}_3=\psi_2=1.072$\\
		$\psi_4=-0.036, O_{\text{s}}(0100, -1), D^{|0100\rangle}_{\{4,9,14,19,24,29,30,31\}}$\\[2pt]
		$\psi_5=-0.036, O_{\text{s}}(0101, -5), D^{|0101\rangle}_{\{0-4,25-31\}}$\\[2pt]
		$\psi_6=0.3, O_{\text{s}}(0110, 6),I^{|0110\rangle}_{6}$\\[2pt]
		$\psi_7=-0.3, O_{\text{s}}(0111, 9),I^{|0111\rangle}_{10}$\\[2pt]
		$\psi_8=0.3, O_{\text{s}}(1000, 11),I^{|1000\rangle}_{12}$\\[2pt]
		$\psi_9=-0.3, O_{\text{s}}(1001, 14),I^{|1001\rangle}_{16}$\\[2pt]
		$\psi_{10}=0.3, O_{\text{s}}(1010, 16),I^{|1010\rangle}_{18}$\\[2pt]
		$\psi_{11}=-0.3, O_{\text{s}}(1011, 19),I^{|1011\rangle}_{22}$\\[2pt]
		$\psi_{12}=0.3, O_{\text{s}}(1100, 21),I^{|1100\rangle}_{24}$\\[2pt]
		$\psi_{13}=-0.3, O_{\text{s}}(1101, 24),I^{|1101\rangle}_{28}$\\
		$\psi_{14}=0$\\
		$\psi_{15}=0$\\
		\hline
	\end{tabular}
	\caption{Block encoding operations (\cref{sec:block_encoding}) for $L^{\dagger}$ of Burgers' equation (\cref{fig:5a_Burgers}), with data vector $v_{\text{data}} = [|\psi_i|]_{i=0}^{15}$. Here, $\tilde{\psi}_2, \tilde{\psi}_3$ denote modified values of $\psi_2 = 1.072, \psi_3=1.144$. Shift operations are given by $O_{\text{s}}(k, diag)$. Here $k$ denote the computational basis state over $|\text{data}\rangle$ qubits and $diag$ denote the diagonal (row minus column). Deletion operations are given by $O_{\text{del}}(k, r) = D^{|k\rangle}_{r}$ (deleting $k^{\text{th}}$ data element from $r^{\text{th}}$ row). Similarly, insert operations are represented as $I^{|k\rangle}_{r}$.}
	\label{table1:Burgers_operations}
\end{table}

The block encoding circuit for operations in \cref{table1:Burgers_operations} is shown in \cref{fig:A3_Block_Encode} which consists of state preparation (PREP, UNPREP), index-mapping oracles (shift, delete, insert), and the coherent permutation (\cref{sec:block_encoding}). We then estimate $\sigma_{\text{min}}$, set $\kappa =8$, and calculate $\phi$ for inverse polynomial of degree 559 for QSVT-based matrix inversion. The solution of Burgers' equation is shown in \cref{fig:5b_Burgers}, where the QSVT implicit scheme is compared with the classical explicit solution of \cref{eq:Burgers_quadratic_ODE} (analogous to the heat equation explicit scheme \cref{eq:heat_explicit}). Results confirm that the proposed pathway (\cref{sec:pathway}) enables quantum linear system methods to solve nonlinear PDEs through Carleman linearization. The analysis of hardware resources and post-selection success probability are reported in \cref{table:Resource_3}.
\begin{table}[ht]
	\centering
	\begin{tabular}{|c|c|}
	\hline
	\multicolumn{2}{|c|}{Burgers' equation}\\
	\hline
	$\text{matrix qubits} (n)$ & $5$\\
	$\text{total qubits}$ & $11$\\
	$\kappa$ & $8$\\
	$d$ & $559$\\
	$p_s$ (analytical) & $0.086517$\\
	$N_{\text{total}}$ & $10^9$\\
	$p_s$ (numerical) & $0.086512$\\
	$\text{BE}_{\text{Hex}}$ & $5687$\\
	$\text{BE}_{\text{Square}}$ & $3892$\\
	$\text{QSVT}_{\text{Hex}}$ & $3.4$M\\
	$\text{QSVT}_{\text{Square}}$ & $2.4$M\\
	\hline
\end{tabular}
	\caption{Estimation of success probability and hardware resources, measured in terms of two-qubit depth, for the Burgers' equation described in \cref{sec:Burgers_equation}. Here $p_s$ (analytical) refers to \cref{eq:p_success_analytical} and $p_s$ (numerical) refers to \cref{eq:p_success}. `M' denotes millions (e.g., 3.4M denotes $3.4 \times 10^6$ two-qubit gate depth). `BE' denotes the block encoding circuit as in \cref{fig:A3_Block_Encode} and `QSVT' denotes the complete QSVT circuit as in \cref{fig:2_QSVT}. `Hex' denotes the heavy-hex lattice topology of the Heron r3 processor (IBM boston), and `Square' denotes the square lattice topology of the Nighthawk r1 processor (IBM miami).}
	\label{table:Resource_3}
\end{table}

\section{Discussion}
In this work, we consolidated key algorithmic and implementation developments and presented a systematic pathway for solving differential equations within the quantum linear systems framework.

A key component of our approach is the use of QSVT in combination with block encoding of sparse matrices for matrix inversion. A notable advantage of this method is its reusability: During QSVT construction, a phase sequence is synthesized to approximate the inverse function over a prescribed spectral interval determined by a design lower bound $\hat{\sigma}_{\text{min}}$, where the approximation is valid for singular values in the range $[\hat{\sigma}_{\text{min}}, 1]$. Once generated, this phase sequence remains applicable to any block encoded matrix $\bar{A} = A/\alpha$ whose smallest singular value satisfies $\sigma_{\text{min}}(\bar{A}) \geq \hat{\sigma}_{\text{min}}$, eliminating the need to recompute phase parameters and thereby reducing preprocessing overhead.

We demonstrated the applicability of this framework through matrix inversion of a complex linear system and through differential equations relevant to computational fluid dynamics. Specifically, we studied the linear heat equation and the Carleman-linearized Burgers’ equation, illustrating how nonlinear dynamics can be embedded into a linear-algebraic formulation suitable for quantum linear solvers.

We analyzed hardware resource requirements, including two-qubit gate depth and post-selection success probability, to estimate the measurement overhead needed to extract physical solutions. Our scaling analysis of the heat equation highlights regimes where classical computation remains feasible and identifies the circuit depths required to achieve potential quantum advantage. This analysis also reveals a major practical limitation: the circuit depth required for polynomial approximations of the inverse function exceeds the coherence-time limits and error thresholds of present-day quantum devices, thereby preventing direct execution on currently available hardware. These findings motivate further investigation into depth-reduction strategies, such as singular value amplification and improved polynomial approximation techniques.

An additional open challenge lies in efficient quantum state sampling and observable estimation, as classical post-processing costs must remain subdominant to preserve theoretical quantum speedups. Addressing these challenges is essential for narrowing the gap between theoretical quantum linear solvers and practical applications.

\section*{Acknowledgements}
This research was funded through the European Union’s Horizon Programme (HORIZONCL4-2021-DIGITALEMERGING-02-10), Grant Agreement 101080085 (QCFD).

\section*{Data Availability}
All data that support the findings of this study are included within the article.

\bibliography{bibliography.bib}

\newpage
\clearpage

\appendix

\begin{center}
    \section*{Appendix}\label{sec:Appendix}
\end{center}
To make the paper self-contained, we include here essential background material. Each section introduces a specific topic with its own notations and conventions.

\section{Quantum Signal Processing}\label{sec:qsp}
Quantum Signal Processing (QSP) is a framework for implementing polynomial transformations on quantum computers. Conceptually, QSP combines two single-qubit rotations: a fixed \emph{signal rotation} $W$ and a variable \emph{signal processing rotation} $S$, tuned to realize the desired polynomial. In this section, we summarize the conventions from \cite{martyn2021grand} used in this work. 
\subsection{$W_X$ Convention for QSP}\label{sec:W_x_convention}
\begin{figure}[t]
	\centering
	\begin{subfigure}[t]{\columnwidth}
		\includegraphics[width=\columnwidth]{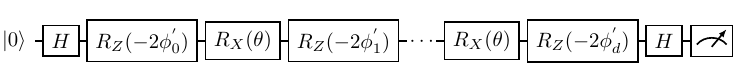}
		\caption{} 
		\label{fig:A1_QSP}
	\end{subfigure}
	\\
	\begin{subfigure}[t]{0.7\columnwidth}
		\includegraphics[width=\columnwidth]{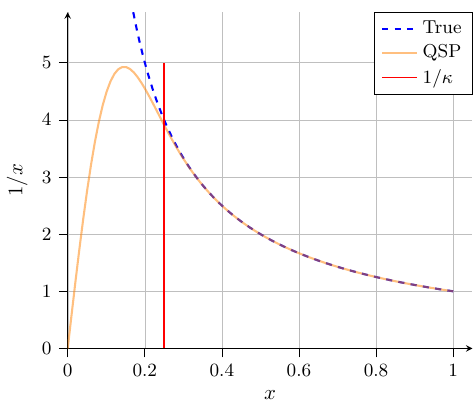}
		\caption{}
		\label{fig:A2_Inverse}
	\end{subfigure}
	\caption{\subref{fig:A1_QSP} Circuit for QSP in the $W_X$ convention \cref{eq:Poly_a}, with $\theta = -2 \cos^{-1}x$. The central dots indicate continuation of the pattern. \subref{fig:A2_Inverse} Approximation of the inverse function using QSP over $x \in [0,1]$. For demonstration, desired inverse values are truncated. The best approximation occurs in the range $x \in [1/\kappa, 1]$.}
	\label{fig:A1_A2}
\end{figure}

In the $W_X$ convention, the signal operator is,
\begin{equation}
	W_X(a) = \begin{bmatrix}
		a & i\sqrt{1-a^2}\\
		i\sqrt{1-a^2} & a
	\end{bmatrix},
\end{equation}
which corresponds to an $R_X(\theta)$ rotation with angle $\theta = -2\cos^{-1}{a}$. The signal processing operator is $S(\phi^{'}) = e^{i \phi^{'}Z} $, with the phase sequence $\vec{\phi^{'}} = (\phi^{'}_0, \phi^{'}_1, \cdots, \phi^{'}_d) \in \mathbb{R}^{d+1}$. The resulting QSP operator is
\begin{equation}\label{eq:W_x_convention}
	U_{W_X}(\vec{\phi^{'}}) = e^{i\phi^{'}_0Z} \prod_{k=1}^d W_X(a) e^{i\phi^{'}_kZ}.
\end{equation}
A real polynomial can be obtained in the basis $M = \{|+\rangle, |-\rangle\}$, leading to the following theorem:
\begin{theorem}\label{theorem_qsp}
	[$(W_X, S_Z, \langle+| \cdot |+\rangle)$-QSP].
	For $a \in [-1,1]$, a phase sequence $\vec{\phi^{\prime}} \in \mathbb{R}^{d+1}$ exists such that
	\begin{equation}\label{eq:Poly_a}
		\textnormal{Poly}(a) = \langle+| U_{W_X}(\vec{\phi^{\prime}}) |+\rangle,
	\end{equation}
	for any real polynomial $\textnormal{Poly} \in \mathbb{R}[a]$, if and only if:
	\begin{enumerate}[label=\roman*., itemsep=0pt]
		\item $\deg(\textnormal{Poly}) \leq d$
		\item $\textnormal{Poly}$ has parity $d \bmod 2$
		\item $|\textnormal{Poly}(a)| \leq 1$ for all $a \in [-1,1]$
	\end{enumerate}
\end{theorem}
The proof is provided in \cite{gilyen2019quantum}. The corresponding circuit is shown in \cref{fig:A1_QSP}. The key task is to compute the phase sequence $\vec{\phi^{\prime}}$ that approximates a target polynomial.

\subsection{Reflection Convention for QSP}\label{sec:reflection_convention}
In the reflection convention, the signal operator is chosen as a reflection:
\begin{equation}
	R(a) = \begin{bmatrix}
		a & \sqrt{1-a^2}\\
		\sqrt{1-a^2} & -a
	\end{bmatrix},
\end{equation}
where $R(a) = XR_Y(\theta)$ with $\theta = 2 \sin^{-1}a$. For a phase sequence $\vec{\phi}$, the QSP operator is
\begin{equation}\label{eq:reflection_convention}
	U_{\text{Reflection}}(\vec{\phi}) = e^{i \phi_0 Z} \prod_{k=1}^{d} R(a) e^{i \phi_k Z}
\end{equation}
The transformation between phase sequences of $W_X$ convention \cref{eq:W_x_convention} $\vec{\phi^{'}}$ and reflection convention \cref{eq:reflection_convention} $\vec{\phi}$ is given by \cite{dong2021efficient, martyn2021grand}, such that,
\begin{equation}
	e^{i \phi_0 Z} \prod_{k=1}^{d} R(a) e^{i \phi_k Z} =  e^{i \phi^{'}_0 Z} \prod_{k=1}^{d} W_X(a) e^{i \phi^{'}_k Z}.
\end{equation}
Note that QSVT requires phase sequences from reflection convention. 

\section{Inverse Function}\label{sec:inverse_function}
In this section, we describe the inverse function approximation using QSP. Specifically, it provides an $\epsilon/2\kappa$ approximation to $(1/2\kappa)(1/x)$ \cite{martyn2021grand}, where $\kappa$ is the condition number of the matrix. The target inversion function is \cite{childs2017quantum, gilyen2019quantum}
\begin{equation}\label{eq:inverse_function}
	f(x) = \frac{1-(1-x^2)^b}{x},
\end{equation}
which is an $\epsilon$-approximation to $1/x$ over $x \in [-1, 1]\setminus[-1/\kappa, 1/\kappa]$.  

Since $f(x)$ is not a polynomial, it is expanded as a linear combination of Chebyshev polynomials over $x\in[-1, 1]$:
\begin{equation}\label{eq:1/x_function}
	\resizebox{\columnwidth}{!}{$
	P^{1/x}_{2\epsilon, \kappa}(x) \!= \!4 \sum\limits_{j=0}^{d}(-1)^j \! \left[ 2^{-2b}\! \sum\limits_{i=j+1}^{b} \!
	\begin{pmatrix}
		2b \\
		b+i
		\end{pmatrix}	\! \right] \! T_{2j+1}(x)$},
\end{equation}
where $b=\kappa^2 \log{\kappa/\epsilon}$, $T_i(x)$ is the Chebyshev polynomial of first kind and order $i$, and 
\[
\resizebox{\columnwidth}{!}{$
d(\epsilon, \kappa) = \left\lceil \sqrt{b(\epsilon, \kappa) \log{[4b(\epsilon, \kappa)/\epsilon]}}\right\rceil = \mathcal{O}[\kappa \log{(\kappa/\epsilon)}]
$}
\]
is the degree of this polynomial. This yields a $2\epsilon$-approximation to $1/x$ over $x\in[-1, 1]\backslash [-1/\kappa, 1/\kappa]$.

The polynomial $P^{1/x}_{2\epsilon, \kappa}(x)$ is then expressed using Laurent polynomials, from which the phase sequence $\vec{\phi^{\prime}}$ in the $W_X$ convention \cref{sec:W_x_convention} is obtained via the \texttt{PyQSP} package \cite{martyn2021grand, chao2020finding, dong2021efficient}. Using these phases, the QSP operator \cref{eq:Poly_a} approximates
\begin{equation}\label{eq:beta}
	\text{Poly}(x) \approx \beta \frac{1}{x},
\end{equation} 
where $\beta$ is a normalization factor, since QSP polynomials are bounded by $1$ on $x \in [-1,1]$ (see \cref{theorem_qsp}). Thus, the approximation to $1/x$ requires dividing by $\beta$.  

As an example, for $\kappa=4$ and $\epsilon=0.1$, the resulting inverse function in the $W_X$ convention is shown in \cref{fig:A2_Inverse}. For use in QSVT, the phase sequence $\vec{\phi^{\prime}}$ is finally converted to the reflection convention $\vec{\phi}$ \cref{sec:reflection_convention}.

\section{Quantum Singular Value Transformation}\label{sec:qsvt}
We have seen that QSP enables polynomial transformations of an input signal $a$ in \cref{theorem_qsp}. QSVT extends this idea by allowing polynomial transformations to be applied directly to the singular values of a block-encoded matrix \cref{eq:U_A_Block_Encode}. Following the conventions in \cite{gilyen2019quantum, martyn2021grand}, we outline QSVT below.

Consider a matrix $A$ scaled by a subnormalization factor $\alpha$, with singular value decomposition (SVD),
\begin{equation}
	A/\alpha = W\Sigma V^{\dag},
\end{equation} 
where $W$ and $V$ are unitary matrices and $\Sigma$ is diagonal with singular values $\Sigma = \{\sigma_k \mid \sigma_k \in [0,1] \, \forall k \}$. There are $r = \text{rank}(A/\alpha)$ nonzero singular values.  

The columns of $W$ and $V$ form orthonormal bases, denoted $\{|w_k\rangle\}$ and $\{|v_k\rangle\}$, corresponding to the left and right singular vectors, respectively. Thus the SVD can be written as,
\begin{equation}\label{eq:A_svd}
	A/\alpha = \sum\limits_{k=1}^{r} \sigma_k |w_k \rangle \langle v_k |.
\end{equation}

From the block-encoding unitary $U_A$ (see \cref{eq:U_A_Block_Encode}), the embedded matrix $A/\alpha$ is obtained as,
\begin{equation}\label{eq:U_A_Projectors}
U_A = \begin{blockarray}{c@{\hskip 1.5ex}cc}
	& \scalebox{0.7}{$\prod$} & & & \\[0.05cm]
	\begin{block}{c@{\hskip 1.5ex}[cc]}
		\raisebox{0.4ex}{\scalebox{0.7}{$\widetilde{\prod}$}} & A/\alpha & *\\
		& * & * \\
	\end{block}
\end{blockarray},
\end{equation}
where $\raisebox{0.3ex}{\scalebox{0.7}{$\widetilde{\prod}$}} = \sum_k |w_k\rangle \langle w_k|$ and $\raisebox{0.3ex}{\scalebox{0.7}{$\prod$}} = \sum_k |v_k\rangle \langle v_k|$ are projectors. In this notation, $A/\alpha = \raisebox{0.4ex}{\scalebox{0.7}{$\widetilde{\prod}$}} \, U_A \, \raisebox{0.4ex}{\scalebox{0.7}{$\prod$}}$. The projector $\raisebox{0.4ex}{\scalebox{0.7}{$\prod$}}$ selects the right singular vectors (columns), while $\raisebox{0.4ex}{\scalebox{0.7}{$\widetilde{\prod}$}}$ selects the left singular vectors (rows). Using these notations, the QSVT can be formulated in theorem as follows,
\begin{theorem}\label{theorem_qsvt}
	For an arbitrary matrix, $A$ block encoded into unitary $U_A$ with subnormalization factor $\alpha$ as, $A/\alpha = \sum_k \sigma_k |w_k \rangle \langle v_k |$, which can be located within $U_A$ using projector-controlled NOT gates $\raisebox{0.3ex}{\scalebox{0.7}{$\prod$}}$ and $\raisebox{0.3ex}{\scalebox{0.7}{$\widetilde{\prod}$}}$ as in \cref{eq:U_A_Projectors}. Then using the projector-controlled phase shift operations $\raisebox{0.3ex}{\scalebox{0.7}{$\prod$}}_\phi$ and $\raisebox{0.3ex}{\scalebox{0.7}{$\widetilde{\prod}$}}_\phi$ (defined as in \cite{martyn2021grand}) for phases $\vec{\phi}$ as in \cref{eq:reflection_convention}, the QSVT operation for odd $d$ is given by,
	\begin{equation}\label{eq:U_phi_QSVT}
		\begin{split}
				\hspace{-0.125cm}
				U_{\vec{\phi}} &=  \raisebox{0.3ex}{\scalebox{0.7}{$\widetilde{\prod}$}}_{\phi_0} U_A \left[ \prod\limits_{k=1}^{(d-1)/2} \raisebox{0.3ex}{\scalebox{0.7}{$\prod$}}_{\phi_{2k-1}} U^{\dag}_A \raisebox{0.3ex}{\scalebox{0.7}{$\widetilde{\prod}$}}_{\phi_{2k}} U_A \right] \raisebox{0.3ex}{\scalebox{0.7}{$\prod$}}_{\phi_{d}} \\
		&= \begin{blockarray}{c@{\hskip 1.5ex}cc}
			& \scalebox{0.7}{$\prod$} & & & \\[0.05cm]
			\begin{block}{c@{\hskip 1.5ex}[cc]}
				\raisebox{0.4ex}{\scalebox{0.7}{$\widetilde{\prod}$}} & \textnormal{Poly}_{(\textnormal{SV})}(A/\alpha) & *\\
				& * & * \\
			\end{block}
		\end{blockarray},
		\end{split}
	\end{equation}
	where $\textnormal{Poly}_{(\textnormal{SV})}(A/\alpha)$ is an odd polynomial of degree $d$ defined as,
	\begin{equation}
		\textnormal{Poly}_{(\textnormal{SV})}(A/\alpha) = \sum_k \textnormal{Poly}(\sigma_k) |w_k \rangle \langle v_k|,
	\end{equation}
	which applies a polynomial transform to the singular values of $A/\alpha$.
	Similarly, for even $d$,
	\begin{equation}
		\begin{split}
			U_{\vec{\phi}} &= \left[ \prod\limits_{k=0}^{d/2} \raisebox{0.3ex}{\scalebox{0.7}{$\prod$}}_{\phi_{2k}} U^{\dag}_A \raisebox{0.3ex}{\scalebox{0.7}{$\widetilde{\prod}$}}_{\phi_{2k+1}} U_A \right] \raisebox{0.3ex}{\scalebox{0.7}{$\prod$}}_{\phi_{d}} \\
			&= \begin{blockarray}{c@{\hskip 1.5ex}cc}
				& \scalebox{0.7}{$\prod$} & & & \\[0.05cm]
				\begin{block}{c@{\hskip 1.5ex}[cc]}
					\raisebox{0.4ex}{\scalebox{0.7}{$\widetilde{\prod}$}} & \textnormal{Poly}_{(\textnormal{SV})}(A/\alpha) & *\\
					& * & * \\
				\end{block}
			\end{blockarray},
		\end{split}
	\end{equation}
	where $\textnormal{Poly}_{(\textnormal{SV})}(A/\alpha)$ is an even polynomial of degree $d$ defined as,
	\begin{equation}
		\textnormal{Poly}_{(\textnormal{SV})}(A/\alpha) = \sum_k \textnormal{Poly}(\sigma_k) |v_k \rangle \langle v_k|,
	\end{equation}
	which is also a polynomial transform to the singular values of $A/\alpha$ but here the input and output vector spaces are both right singular vectors.
\end{theorem}
The proof of this theorem, along with the general block-encoding framework, is given in \cite{gilyen2019quantum}. A key advantage of QSVT is that it applies polynomial transformations directly to the singular values of a matrix, without requiring their explicit computation or an SVD.

\section{Finite Difference Methods for Heat Equation}\label{sec:FDE_heat}
Consider the heat equation $\frac{\partial u}{\partial t} = \nu \frac{\partial^2 u}{\partial x^2}$, where $\nu > 0$ is the thermal diffusivity. The discrete variables are $t_n = n \Delta t, \, x_i = i \Delta x, \, u_i^n = u(x_i, t_n)$. Using a forward difference in time and central difference in space, the explicit scheme is
\begin{equation}
	\frac{u_i^{n+1} - u_i^{n}}{\Delta t} = \nu \frac{u_{i+1}^n - 2u_i^n + u_{i-1}^n}{\Delta x^2} + O(\Delta t, \Delta x^2),
\end{equation}
where $O(\Delta t, \Delta x^2)$ denotes the truncation error. Rearranging gives
\begin{equation}\label{eq:heat_explicit}
	u^{n+1}_i = \lambda u_{i+1}^n + (1 - 2 \lambda) u_i^n + \lambda u_{i-1}^n,
\end{equation}
with $\lambda = \nu \frac{\Delta t}{\Delta x^2}$. In vector form, this is $\vec{u}^{n+1} = A \vec{u}^n + \vec{f}^{\text{BC}}$, where $\vec{f}^{\text{BC}}$ accounts for boundary conditions. This scheme avoids matrix inversion but is conditionally stable, with stability governed by von Neumann analysis \cite{leveque2007finite, ames2014numerical}:
\begin{equation}\label{eq:heat_von_Neumann}
	\Delta t_{\text{exp}}^{\text{max}} \leq \frac{\Delta x^2}{2\nu}.
\end{equation}
For $\Delta t > \Delta t_{\text{exp}}^{\text{max}}$, the explicit scheme diverges, while the implicit method remains stable for any $\Delta t$, making it preferable for larger time steps.  

The implicit discretization is
\begin{equation}\label{eq:heat_implicit_discretize}
	\frac{u_i^{n+1} - u_i^{n}}{\Delta t} = \nu \frac{u_{i+1}^{n+1} - 2u_i^{n+1} + u_{i-1}^{n+1}}{\Delta x^2} + O(\Delta t, \Delta x^2),
\end{equation}
which rearranges to
\begin{equation}\label{eq:heat_implicit}
	- \lambda u_{i-1}^{n+1} + (1 + 2\lambda) u_i^{n+1} - \lambda u_{i+1}^{n+1} = u_i^n.
\end{equation}
This leads to the linear system
\begin{equation}\label{eq:heat_LSE}
	A \vec{u}^{n+1} = \vec{b}^n,
\end{equation}
where $\vec{b}^n = \vec{u}^n + \vec{f}^{\text{BC}}$. Such systems can be efficiently solved using QSVT \cref{sec:matrix_inversion_by_qsvt}.  

For boundary conditions, let the solution domain be $x \in [0, L]$ with grid points $\{u_i\}_{i=0}^N$. We impose a Dirichlet condition on the left boundary, $u_0 = D_B^l$, and a Neumann condition on the right, $\partial u/\partial x|_{x=L} = N_B^r$, approximated using central difference as
\begin{equation}
	\frac{u_{N+1} - u_{N-1}}{2\Delta x} = N_B^r,
\end{equation}
where the ghost point method gives $u_{N+1} = u_{N-1} + 2N_B^r \Delta x$. Substituting these boundary conditions into \cref{eq:heat_implicit} yields the linear system \cref{eq:heat_LSE}, where $\vec{u}^{n+1} = [u_1^{n+1}, u_2^{n+1}, \ldots, u_N^{n+1}]^T, \quad 
\vec{b}^n = [u_1^n + \lambda D_B^l, u_2^n, \ldots, u_N^n + 2\lambda N_B^r \Delta x]^T,$
and the coefficient matrix is
\begin{equation}\label{eq:heat_A}
	A = \begingroup
	\setlength{\arraycolsep}{2.3pt}
	\renewcommand{\arraystretch}{1.2}
	\begin{bmatrix}
		1 + 2\lambda & -\lambda \\
		-\lambda & 1 + 2\lambda & -\lambda \\
		& \ddots & \ddots & \ddots \\
		&& -\lambda & 1 + 2\lambda & -\lambda \\
		&&& -2\lambda & 1 + 2\lambda
	\end{bmatrix}.
	\endgroup
\end{equation}

The matrix $A$ in \cref{eq:heat_A} is a near-Toeplitz tridiagonal matrix. Following the spectral analysis in \cite{strikwerda2004finite,leveque2007finite,morton2005numerical}, its eigenvalues are given by
\begin{equation}
	\lambda_k = 1 + 4\lambda \sin^2\left({\frac{(2k-1)\pi}{4N}}\right),
\end{equation}
where $k=1, 2, \cdots, N$. The smallest eigenvalue corresponds to $k=1$: $\lambda_{\text{min}} = 1 + 4\lambda \sin^2\left({\frac{\pi}{4N}}\right)$, and the largest eigenvalue corresponds to $k=N$: $\lambda_{\text{max}} = 1 + 4\lambda \sin^2\left({\frac{(2N-1)\pi}{4N}}\right) = 1 + 4\lambda \cos^2\left({\frac{\pi}{4N}}\right)$. 

Since $A$ is real and only weakly non-symmetric (with symmetry broken by a single boundary entry), its singular values are well approximated by the absolute values of its eigenvalues. Moreover, all eigenvalues are positive. Hence, we approximate $\sigma_{\max}(A) \approx \lambda_{\max}$ and $\sigma_{\min}(A) \approx \lambda_{\min}$.

In the scaling analysis presented in \cref{fig:4c_Heat,fig:4d_Heat,fig:4e_Heat}, we consider the subnormalization factor $\alpha = \sigma_{\max}$ so that the scaled matrix satisfies $\sigma_{\max} = 1$ (see \cref{eq:U_A_Block_Encode}). The minimum singular value of the scaled matrix is therefore approximated by
\begin{equation}\label{eq:sigma_min_scaled}
	\sigma_{\min}^{\text{scaled}} \approx \frac{\lambda_{\min}}{\lambda_{\max}} = \frac{1 + 4\lambda \sin^2\left(\frac{\pi}{4N}\right)}{1 + 4\lambda \cos^2\left(\frac{\pi}{4N}\right)}.
\end{equation}
Since $\Delta x = 2^{-q}$, where $q$ is the number of qubits representing the spatial grid, the number of interior grid points is $N = 2^{q}$. Consequently,
\begin{equation}\label{eq:lambda}
	\lambda = \frac{\nu\Delta t}{(\Delta x)^2} = \nu\Delta t \; 4^{q}.
\end{equation}
Substituting \cref{eq:lambda} into \eqref{eq:sigma_min_scaled} yields the explicit expression in terms of $q$:
\begin{equation}\label{eq:sigma_min_estimation}
	\sigma_{\min}^{\text{scaled}} \approx \frac{\lambda_{\min}}{\lambda_{\max}} = \frac{1 + 4\nu \Delta t \; 4^{q} \sin^2\left(\frac{\pi}{4 \cdot 2^q}\right)}{1 + 4\nu \Delta t \; 4^q \cos^2\left(\frac{\pi}{4 \cdot 2^q}\right)}.
\end{equation}
This formula can be evaluated directly for any $q$ in $O(1)$ time, allowing estimation of $\sigma_{\min}$ for values such as $q=50$ without explicitly constructing the $2^{50}\times 2^{50}$ matrix. Furthermore, the analytical expression in \cref{eq:sigma_min_estimation} agrees with the extrapolated curve-fitting results shown in \cref{fig:4d_Heat,fig:4e_Heat}. 

For large $q$, we derive an asymptotic approximation by defining $\varepsilon = \pi/(4\cdot 2^q) \ll 1$. Using $\sin^2\varepsilon \approx \varepsilon^2 = \pi^2/(16\cdot 4^q)$ and $\cos^2\varepsilon \approx 1 - \varepsilon^2$, and noting that $4\nu\Delta t \cdot 4^q \gg 1$ for large $q$, we obtain
\begin{equation}\label{eq:sigma_min_approx}
	\sigma_{\min}^{\text{scaled}}(q) \approx \frac{4\nu\Delta t \cdot 4^q \cdot \frac{\pi^2}{16\cdot 4^q}}{4\nu\Delta t \cdot 4^q} = \frac{\pi^2}{16} \cdot 4^{-q}.
\end{equation}
Thus, the minimum singular value decays exponentially as $O(4^{-q})$. Consequently, the condition number of the scaled matrix behaves as
\begin{equation}
	\kappa = \frac{1}{\sigma_{\min}^{\text{scaled}}(q)} \sim \frac{16}{\pi^2} \cdot 4^q.
\end{equation}

Since the QSVT algorithm scales as $O(\kappa \log(1/\epsilon))$ \cite{gilyen2019quantum,martyn2021grand}, this exponential growth of the condition number with $q$ presents a fundamental challenge when solving the heat equation at high spatial resolution. Conversely, for fixed $q$, increasing the time steps $\Delta t$ increase $\lambda$ and therefore increases $\sigma_{\min}^{\text{scaled}}$, improving the condition number, consistent with the unconditional stability of the implicit scheme.

\section{Carleman Linearization}\label{sec:Carleman_Linearization}
Consider the initial value problem defined by an $n$-dimensional quadratic ordinary differential equation (ODE) \cite{liu2021efficient}:
\begin{equation}\label{eq:quadratic_ode}
	\frac{\text{d}u}{\text{d}t} = F_2 u^{\otimes2} + F_1 u + F_0(t), \quad u(0) = u_{\text{in}},
\end{equation}
where $u = [u_1, \ldots, u_n]^T \in \mathbb{R}^n$, and 
$u^{\otimes2} = [u_1^2, \, u_1u_2, \ldots, u_1u_n, \, u_2u_1, \ldots, u_n u_{n-1}, \, u_n^2]^T \in \mathbb{R}^{n^2}$.  
Each component $u_j = u_j(t)$ is defined on the interval $[0, T]$ for $j \in [n] = \{1, \ldots, n\}$. The coefficient matrices are $F_2 \in \mathbb{R}^{n \times n^2}$ and $F_1 \in \mathbb{R}^{n \times n}$, both time-independent, while the inhomogeneity $F_0(t) \in \mathbb{R}^n$ is a $C^1$ continuous function of $t$.  
\begin{figure*}[t]
	\centering
	\includegraphics[width=0.9\textwidth]{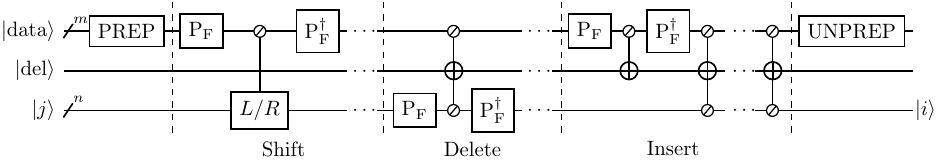}
	\caption{Circuit for block encoding via coherent permutation \cite{setty2025block}. 
		The circuit is initialized with $m$ data qubits ($|\text{data}\rangle$), one delete qubit ($|\text{del}\rangle$), and $n$ matrix index qubits ($|j\rangle$). 
		It employs PREP and UNPREP operators for state preparation, along with combined shift, delete, and insert operations. 
		Coherent permutation is handled by $\text{P}_{\text{F}}$ operators for a given fixed indices $F$. Dots indicate repeating structures.}
	\label{fig:A3_Block_Encode}
\end{figure*}
Carleman linearization is a technique for transforming a finite-dimensional nonlinear system into an infinite hierarchy of coupled linear differential equations \cite{carleman1932application,kowalski1991nonlinear,forets2017explicit}. Truncating this hierarchy at order $N$ yields a finite linear system. For the quadratic ODE system \cref{eq:quadratic_ode}, the Carleman procedure (see \cite{liu2021efficient}) produces the corresponding system of linear ODEs as follows:
\begin{equation}\label{eq:Carleman_equation}
	\frac{\text{d}\vec{y}}{\text{d}t} = A(t)\vec{y} + b(t), \hspace{0.4cm} \vec{y}(0) = \vec{y}_{\text{in}},
\end{equation}
where $\vec{y} = [\hat{y}_1, \hat{y}_2, \cdots, \hat{y}_{N-1}, \hat{y}_N]^T, b(t)=[F_0(t), 0, \cdots, 0]^T$ and matrix $A$ is given as,
\begin{equation}\label{eq:Carleman_A}
	A =  \begingroup
	\setlength{\arraycolsep}{2.3pt}
	\renewcommand{\arraystretch}{1.3}
	\begin{bmatrix}
		A_1^1 & A_2^1\\
		A_1^2 & A_2^2 & A_3^2\\
		& \ddots & \ddots & \ddots\\
		&& A_{N-2}^{N-1} & A_{N-1}^{N-1} & A_{N}^{N-1}\\
		&&& A_N^{N-1} & A_N^N
	\end{bmatrix}.
	\endgroup
\end{equation}
Here $\hat{y}_j=u^{\otimes j} \in \mathbb{R}^{n^j}, \vec{y}_{\text{in}} = [u_{\text{in}}; u_{\text{in}}^{\otimes 2}; \cdots ; u_{\text{in}}^{\otimes N}],$ and $A^j_{j+1} \in \mathbb{R}^{n^j \times n^{j+1}}, A_j^j \in \mathbb{R}^{n^j \times n^j}, A_{j-1}^j \in \mathbb{R}^{n^j \times n^{j-1}}$ for $j \in [N]$ satisfying
\begin{equation}
	\resizebox{\columnwidth}{!}{$
	A_{j+1}^j = F_2 \otimes I^{\otimes j-1} + I \otimes F_2 \otimes I^{\otimes j-2} + \cdots + I^{\otimes j-1} \otimes F_2,$}
\end{equation}
\begin{equation}
	\resizebox{\columnwidth}{!}{$
		A_{j}^j = F_1 \otimes I^{\otimes j-1} + I \otimes F_1 \otimes I^{\otimes j-2} + \cdots + I^{\otimes j-1} \otimes F_1,$}
\end{equation}
\begin{equation}
	\resizebox{\columnwidth}{!}{$
		A_{j-1}^{j} \!= \! F_0(t) \! \otimes \! I^{\otimes j-1}\! +\! I \! \otimes \! F_0(t) \! \otimes \! I^{\otimes j-2} \! + \! \cdots \! + \! I^{\otimes j-1} \! \otimes \! F_0(t).$}
\end{equation}
The dimension of \cref{eq:Carleman_A} is,
\begin{equation}\label{eq:Dim_A}
	\text{Dim}(A) = n + n^2 + \cdots + n^N = \frac{n^{N+1}-n}{n-1} = O(n^N).
\end{equation}
The Carleman-linearized system \cref{eq:Carleman_equation} can be expressed as a linear system of equations using the implicit finite difference method from \cref{sec:FDE_heat}. For discrete time steps $t_k = k \Delta t$, the scheme takes the form
\begin{equation}\label{eq:Carleman_linear_system}
	\begin{split}
		\frac{\vec{y}^{\,k+1} - \vec{y}^{\,k}}{\Delta t} &= A(t)\,\vec{y}^{\,k+1} + b(t),\\
		(I - A(t)\Delta t)\,\vec{y}^{\,k+1} &= \vec{y}^{\,k} + b(t)\Delta t,\\
		L\,\vec{y}^{\,k+1} &= B.
	\end{split}
\end{equation}

\section{Linearization of Burgers' Equation}\label{sec:Linearization_Burgers}
Consider the one-dimensional Burgers' equation,  
\begin{equation}
	\frac{\partial u}{\partial t} + u \frac{\partial u}{\partial x} = \nu \frac{\partial^2 u}{\partial x^2},
\end{equation}
where $\nu \,(\text{const.} \geq 0)$ is the diffusion coefficient, $x \in [0, 1]$, and $t \in [0, T]$. The discrete variables are defined as $t_k = k \Delta t, x_i = i \Delta x, u_i^k = u(x_i, t_k)$ with $\{u_i\}_{i=0}^{S}$. Applying a central difference discretization in space gives
\begin{equation}\label{eq:Burgers_quadratic_ODE}
	\resizebox{\columnwidth}{!}{$
		\frac{\text{d} u}{\text{d} t} 
		= \frac{\nu}{\Delta x^2}(u_{i+1} - 2u_i + u_{i-1}) 
		- \frac{1}{2\Delta x}\, u_i (u_{i+1} - u_{i-1}).$}
\end{equation}
This form directly matches the quadratic ODE representation in \cref{eq:quadratic_ode}. For the test problem in \cref{sec:Burgers_equation}, we impose Dirichlet boundary conditions $u(0, t) = u_0 = u(1, t) = u_{S-1} = 0$. Defining $\lambda_1 = \tfrac{\nu}{\Delta x^2}$ and $\lambda_2 = -\tfrac{1}{2\Delta x}$, the system matrices in \cref{eq:quadratic_ode} are
\begin{equation}
	F_1 = \lambda_1 
	 \begingroup
	\setlength{\arraycolsep}{2.3pt}
	\renewcommand{\arraystretch}{1.1}
	\begin{bmatrix}
		-2 & 1\\
		1 & -2 & 1\\
		& \ddots & \ddots & \ddots\\
		&&1 & -2 & 1\\
		&&& 1 &-2
	\end{bmatrix}_{(S-2, S-2)},
	\endgroup
\end{equation}

\begin{equation}
	F_2 \!= \! \lambda_2\! \\
	\begingroup
	\setlength{\arraycolsep}{2pt}
	\renewcommand{\arraystretch}{1.1}
	\begin{bmatrix}
		0&-1\\
		&&\cdots&1&0&-1\\
		&&&&&\vdots\\
		&&&&&&1&0
	\end{bmatrix}_{(S-2, (S-2)^2)},
	\endgroup
\end{equation}
with $F_0(t)=0$. For a truncation order $N=2$, the Carleman matrix \cref{eq:Carleman_A} becomes
\begin{equation}\label{eq:A_truncation}
	A = \begin{bmatrix}
		A_1^1 & A_2^1\\
		A_1^2 & A_2^2
	\end{bmatrix},
\end{equation}
of size $(S - 2 + (S - 2)^2, S - 2 + (S - 2)^2)$. Since $F_0(t)=0$, we have $A_1^2=0$. This can be formulated further as linear system using implicit scheme \cref{eq:Carleman_linear_system} that can be solved using QSVT \cref{sec:matrix_inversion_by_qsvt}.

\begin{figure*}[t]
	\centering
	\begin{subfigure}[t]{0.37\textwidth}
		\includegraphics[width=\textwidth]{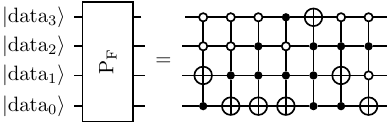}
		\caption{}
		\label{fig:A4_Permutation}
	\end{subfigure}
	\hspace{1cm}
	\begin{subfigure}[t]{0.33\textwidth}
		\includegraphics[width=\textwidth]{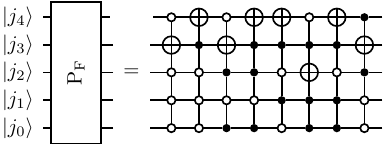}
		\caption{}
		\label{fig:A5_Permutation}
	\end{subfigure}
	\label{fig:A4_A5_Permutation}
	\caption{\subref{fig:A4_Permutation} Coherent permutation on data qubits (\cref{eq:right_permute_mapping}). \subref{fig:A5_Permutation} Coherent permutation on matrix qubits (\cref{eq:left_permute_mapping}).}
\end{figure*}

\section{Block Encoding of Sparse Matrices}\label{sec:block_encoding}
In this section, we present the final circuit for block encoding of sparse matrices using coherent permutation \cite{setty2025block}. The approach employs PREP/UNPREP operators in conjunction with shift (left $L$, right $R$), delete, and insert operators. The coherent permutation mechanism plays a central role in reducing the control overhead of multi-controlled X (MCX) gates and achieving nearest-neighbor connectivity, resulting in hardware-efficient circuit designs. The overall construction is depicted in \cref{fig:A3_Block_Encode}, where $\text{P}_{\text{F}}$ represent permutation with fixed indices set $F$. Further implementation details are provided in \cite{setty2025block}.

To shift $k^{\text{th}}$ data element to diagonal $diag$ (row minus column), we first find the corresponding shifts as powers of two, $|diag| = \sum_{l=0}^{n-1}b_l 2^l, b_l \in \{0,1\}.$ For an integer $|diag|$ in binary $(b_{n-1}\cdots b_0)_2$, we define a set of 1-bit positions  as $B(|d|) = \{l | b_l=1\}$. Then the shift operator is given by,
\begin{equation}\label{eq:O_shift}
	O_{\text{s}}(k, diag) = \begin{cases}
		\prod_{b\in B(|diag|)}L(k, b) \; \text{if} \; diag>0,\\
		\prod_{b\in B(|diag|)}R(k, b) \; \text{if} \; diag<0.
	\end{cases}
\end{equation}

For the block encoding of the Carleman-linearized matrix of the Burgers’ equation discussed in \cref{sec:Burgers_equation}, the corresponding operations are listed in \cref{table1:Burgers_operations}. Note that two distinct diagonal entries, $(\psi_2, \psi_3)$, occur along the main diagonal. Within the block-encoding framework, these appear as a combined term $\psi_2 + \psi_3$ (see \cite{setty2025block}). To address this case, we modify block-encoding operations as follows: (i) data vector: $\tilde{\psi}_2 = \psi_3 - \psi_2, \tilde{\psi}_3 = \psi_2$; (ii) delete operations: $\tilde{\psi}_2: D^{\tilde{\psi}_2}_{\{0-4\}}, \tilde{\psi}_3: \text{Nothing}$. The resulting subnormalization factor within $\alpha$ becomes $|\tilde{\psi}_2| + |\tilde{\psi}_3|$, which is advantageous when $|\psi_3 - \psi_2| < |\psi_3|$. Next, we identify the common shift operators, summarized in \cref{table:Common_shift}, while the delete and insert operations remain as defined in \cref{table1:Burgers_operations}.

\begin{table}[ht]
	\centering
	\begin{tabular}{| l | l |}
		\hline
		\multicolumn{2}{|c|}{Common shift operators}\\
		\hline
		$L(k, 0)$ &$(\psi_0, 0000), (\psi_1, 0001), (\psi_7, 0111), $\\
		&$(\psi_8, 1000), (\psi_{11}, 1011), (\psi_{12}, 1100),$\\
		&$(\psi_{14}, 1110), (\psi_{15}, 1111)$\\
		$L(k, 1)$ &$(\psi_6, 0110), (\psi_8, 1000), (\psi_9, 1001)$\\
		&$(\psi_{11}, 1011)$\\
		$L(k, 2)$ & $(\psi_0, 0000), (\psi_6, 0110), (\psi_9, 1001)$\\
		&$(\psi_{12}, 1100)$\\
		$L(k, 3)$ & $(\psi_7, 0111), (\psi_8, 1000), (\psi_9, 1001)$\\
		&$({13}, 1101)$\\
		$L(k, 4)$ & $(\psi_{10}, 1010), (\psi_{11}, 1011), (\psi_{12}, 1100),$\\
		&$(\psi_{13}, 1101)$\\
		$R(k, 0)$ & $(\psi_4, 0100), (\psi_5, 0101)$\\
		$R(k, 2)$ & $(\psi_5, 0101)$\\
		\hline
	\end{tabular}
	\caption{Common shift operations of the data elements \cref{table1:Burgers_operations}. Note that the $L(k, 0)$ shift involves six data elements; however, zeros in $\psi_{14}$ and $\psi_{15}$ can be exploited to extend this to eight shifts.}
	\label{table:Common_shift}
\end{table}
We now determine the amplitude permutation required for combined operations. For demonstration, we consider the shift $L(k, 0)$ from \cref{table:Common_shift} for which we determine $F=\{|\text{data}_0\rangle\}$ and apply the permutation operator $\text{P}_\text{F}$ to reduce control overhead of MCX gates and achieve nearest neighbor connectivity. Following the block encoding protocol \cite{setty2025block}, the permutation of amplitudes is given by the following walk operators:
\begin{equation}\label{eq:right_permute_mapping}
	\begin{aligned}
		\psi_0\hspace{0.2cm}&0000\\
		\psi_1\hspace{0.2cm}&0001&\rightarrow0011&\rightarrow0010\\
		\psi_7\hspace{0.2cm}&0111&\rightarrow0110\\
		\psi_8\hspace{0.2cm}&1000\\
		\psi_{11}\hspace{0.2cm}&1011&\rightarrow1010\\
		\psi_{12}\hspace{0.2cm}&1100\\
		\psi_{14}\hspace{0.2cm}&1110\\
		\psi_{15}\hspace{0.2cm}&1111&\rightarrow0111&\rightarrow0101&\rightarrow0100
	\end{aligned}.
\end{equation}
The MCX gates implementing the mapping in \cref{eq:right_permute_mapping} is illustrated in \cref{fig:A4_Permutation}.

Considering the combined delete $D^{|0001\rangle}_{\{0,5,10,15,20,25,30,31\}}$ (refer \cref{table1:Burgers_operations}), we determine $F = \{|j_4 j_3\rangle\}$ and apply the permutation operator $\text{P}_\text{F}$. The permutation of amplitudes is given by the following walk operators:
\begin{equation}\label{eq:left_permute_mapping}
	\begin{aligned}
		0\hspace{0.2cm}&00000&\rightarrow01000&\rightarrow11000\\
		5\hspace{0.2cm}&00101&\rightarrow01101&\rightarrow11101\\
		10\hspace{0.2cm}&01010&\rightarrow11010\\
		15\hspace{0.2cm}&01111&\rightarrow01011&\rightarrow11011\\
		20\hspace{0.2cm}&10100&\rightarrow11100\\
		25\hspace{0.2cm}&11001\\
		30\hspace{0.2cm}&11110\\
		31\hspace{0.2cm}&11111
	\end{aligned}.
\end{equation}
The MCX gates corresponding to the mapping in \cref{eq:left_permute_mapping} is illustrated in \cref{fig:A5_Permutation}. The complete block-encoding circuit for the Carleman-linearized matrix is shown in \cref{fig:A3_Block_Encode}, and this circuit is embedded within the QSVT framework (\cref{fig:2_QSVT}) to solve the Burgers’ equation described in \cref{sec:Burgers_equation}.
\end{document}